\documentclass[%
 reprint,
 amsmath,amssymb,
 aps,
]{revtex4-1}

\usepackage{graphicx}% Include figure files
\usepackage{dcolumn}% Align table columns on decimal point
\usepackage{bm}% bold math
\begin{document}

\preprint{APS/123-QED}

\title{Correlation of Co-Located Hydrogen Masers}% Force line breaks with \\
%\thanks{A footnote to the article title}%

\author{Y.C. Guo,$^{1}$ B. Wang,$^{1,*}$ H.W. Si,$^1$ Z.W. Cai,$^3$ A.M. Zhang,$^4$ X. Zhu,$^5$ J. Yang,$^5$ K.M. Feng,$^5$ C.H. Han,$^3$ T.C. Li,$^4$ and L.J. Wang$^{1,2,*}$}

\affiliation{
$^1$State Key Laboratory of Precision Measurement Technology and Instruments, Department of Precision Instrument, Tsinghua University, Beijing 100084, China\\
$^2$Department of Physics, Tsinghua University, Beijing 100084, China\\
$^3$Beijing Satellite Navigation Center, Beijing 100094, China\\
$^4$National Institute of Metrology, Beijing 100013, China\\
$^5$Beijing Institute of Radio Metrology and Measurement, Beijing 100854, China\\
$^*$Corresponding author: bo.wang@tsinghua.edu.cn, lwan@tsinghua.edu.cn}

\date{\today}% It is always \today, today,
             %  but any date may be explicitly specified

\begin{abstract}
The correlation of co-located hydrogen masers (H-masers) is difficult to measure because their common-mode noise induced by the environment will be cancelled out during the comparison measurement. With the development of fibre-based high-precision time and frequency transfer technique, the correlation of co-located hydrogen masers can be directly measured with the help of remote H-masers. Recently, a fiber-based frequency synchronization network was constructed in the Beijing region by connecting 5 H-masers from 4 institutions. The correlation coefficient of atomic clocks is defined and the correlation between two co-located H-masers is measured using both experimental and simulative methods. The results show that the correlation is not prominent until the averaging time is larger than $\sim10^3$s; then, the coefficient grows rapidly for averaging times ranging from $\sim10^3$s to $\sim10^5$s and decreases beyond $\sim10^5$s up to 5 days.
\begin{description}
%\item[Usage]
%Secondary publications and information retrieval purposes.
\item[PACS numbers]
06.30.Ft, 06.20.-f, 06.90.+v
%\item[Structure]
%You may use the \texttt{description} environment to structure your abstract;
%use the optional argument of the \verb+\item+ command to give the category of each item.
\end{description}
\end{abstract}

\pacs{Valid PACS appear here}% PACS, the Physics and Astronomy
                             % Classification Scheme.
%\keywords{Suggested keywords}%Use showkeys class option if keyword
                              %display desired
\maketitle

%\tableofcontents

\section{\label{intro}INTRODUCTION}

Hydrogen masers (H-masers), owing to their excellent frequency stability, are commonly used for precise time-keeping. There exists a certain correlation among hydrogen masers at the same location since the frequency of an H-maser is affected by environmental factors such as temperature, pressure, humidity, magnetic fields, line voltage, and ground loops~\cite{1,2}. This correlation is undesired in many cases. For example, when a cluster of co-located H-masers are constantly compared to generate a weighted-average time, the correlation may lead to inaccurate predictions and weighting of the clocks, which will eventually affect the time-keeping performance~\cite{3}. Some laboratories place H-masers in separate environmental chambers to avoid common-mode frequency fluctuations. However, some environmental factors are too difficult or expensive to control. Consequently, in practical applications, the correlation still exists. Therefore, it is of great importance to accurately determine the correlation of co-located H-masers.

This correlation has always been difficult to measure. In many laboratories, the only frequency source with a short-term frequency stability comparable to that of an H-maser is another H-maser. The correlation cannot be detected because the environment-induced common-mode noise is cancelled when the clocks are measured against each other. Therefore, it is necessary to compare the clocks with other frequency sources under different environmental conditions. In the comparison of remote H-masers, the technique of long-distance time and frequency transfer plays an important role. Recently, with the development of fibre-based time and frequency synchronization~\cite{4,5,6,7,8,9,10,11,12,13,14}, high-precision comparisons of distant clocks have been achieved~\cite{15,16,17,18,19,20}. Correlation measurement between co-located clocks becomes feasible with the help of this technology.

In this paper, we introduce a method for measuring the correlation between co-located H-masers and define the coefficient of clock correlation. Using the frequency synchronization fibre network that we constructed in Beijing over the past few years~\cite{10,21,22}, we measured the correlation of two co-located H-masers by comparing them with three other H-masers separately located. In addition, we built a clock model to simulate the specifications of H-masers under the influence of primary environmental factors and calculated the correlation between two H-masers in the same environment. By comparing the experimental and simulative results, we demonstrate the validity of the correlation measurement and analyse the causes of the correlation properties.

\section{\label{coefficient}coefficient of clock correlation}

The frequency stability of an atomic clock is often represented by the Allan variance:
\begin{equation}
\sigma^2(\tau)=\frac{1}{2M}\sum_{i=1}^{M}[\Delta y^i(\tau)]^2,
\label{eq1}
\end{equation}
where $\Delta y^i(\tau)$ is the difference between the $(i+1)^{th}$ and $i^{th}$ fractional frequencies averaged over the measurement interval $\tau$: $\Delta y^i(\tau)=\bar{y}^{i+1}(\tau)-\bar{y}^i(\tau)$. $M$ is the number of frequency difference samples.

In practice, the frequency stability of an atomic clock can only be measured through comparison with other clocks. The result of such measurement includes not only the instability of the target clock but also that of the reference clock. The Allan variance of two clocks, A and B, can be expressed as follows (drift removed):
\begin{equation}
\sigma^2_{AB}(\tau)=\sigma^2_{A}(\tau)+\sigma^2_{B}(\tau)-C_{AB}(\tau),
\label{eq2}
\end{equation}
where $\sigma_{AB}^2$ is the relative frequency stability, and $\sigma_A^2$ and $\sigma_B^2$ are the individual stabilities of clocks A and B, respectively. The last term $C_{AB}=\frac{1}{M}\sum_{i=1}^M[\Delta y_A^i(\tau)\Delta y_B^i(\tau)]$, is the correlation between the two clocks. Using the traditional stability measurement method, we cannot measure or solve for this term independently. In most cases, we have to assume that there is no correlation between the two clocks or that the correlation is negligibly small, such that the last term vanishes. To obtain the frequency stability of each clock individually,$\frac{1}{2}\sigma_{AB}^2$ is frequently taken as both $\sigma_{A}^2$ and $\sigma_{B}^2$, or the three-cornered-hat method is applied when there are three or more clocks~\cite{23,24,25,26}.

To investigate the correlation term, consider 4 clocks, A, B, C and D, of which A and B are co-located and C and D are placed at two different locations. Therefore, we assume that, except for the clock pair A and B, the correlation terms between other pairs of clocks (A and C, A and D, B and C, B and D, C and D) are negligible. In this case, the frequency stability equations can be expressed as follows:
\begin{equation}
\sigma^2_{AB}(\tau)=\sigma^2_{A}(\tau)+\sigma^2_{B}(\tau)-C_{AB}(\tau),
\label{eq3}
\end{equation}
\begin{equation}
\sigma^2_{AC}(\tau)=\sigma^2_{A}(\tau)+\sigma^2_{C}(\tau),
\label{eq4}
\end{equation}
\begin{equation}
\sigma^2_{AD}(\tau)=\sigma^2_{A}(\tau)+\sigma^2_{D}(\tau),
\label{eq5}
\end{equation}
\begin{equation}
\sigma^2_{BC}(\tau)=\sigma^2_{B}(\tau)+\sigma^2_{C}(\tau),
\label{eq6}
\end{equation}
\begin{equation}
\sigma^2_{BD}(\tau)=\sigma^2_{B}(\tau)+\sigma^2_{D}(\tau),
\label{eq7}
\end{equation}
\begin{equation}
\sigma^2_{CD}(\tau)=\sigma^2_{C}(\tau)+\sigma^2_{D}(\tau).
\label{eq8}
\end{equation}
Using the three-cornered-hat method for clocks A, C, and D, we obtain the individual frequency stability of clock A from Eq.~(\ref{eq4}),~(\ref{eq5}) and~(\ref{eq8}):
\begin{equation}
\sigma^2_{A}(\tau)=\frac{1}{2}[\sigma^2_{AC}(\tau)+\sigma^2_{AD}(\tau)-\sigma^2_{CD}(\tau)].
\label{eq9}
\end{equation}
In the same way, with clocks B, C, and D, the individual frequency stability of clock B is calculated from Eq.~(\ref{eq6}),~(\ref{eq7}) and~(\ref{eq8}) as follows:
\begin{equation}
\sigma^2_{B}(\tau)=\frac{1}{2}[\sigma^2_{BC}(\tau)+\sigma^2_{BD}(\tau)-\sigma^2_{CD}(\tau)].
\label{eq10}
\end{equation}
Consequently, we calculate the correlation term between clock A and B:
\begin{equation}
C_{AB}(\tau)=\sigma^2_{A}(\tau)+\sigma^2_{B}(\tau)-\sigma^2_{AB}(\tau).
\label{eq11}
\end{equation}

We define the coefficient of clock correlation between clock A and B as half of the correlation term divided by the product of their individual Allan deviations. We use $\gamma_{AB}(\tau)$ to represent this correlation coefficient:
\begin{equation}
\gamma_{AB}(\tau)=\frac{C_{AB}(\tau)}{2\sigma_{A}(\tau)\sigma_{B}(\tau)}.
\label{eq12}
\end{equation}
The absolute value of the coefficient $\gamma_{AB}(\tau)$ is less than or equal to 1. The closer $\gamma_{AB}(\tau)$ is to zero, the weaker the correlation is. If the coefficient $\gamma_{AB}(\tau)$ equals 1 or $-1$, the frequency fluctuations of the two clocks are linearly related. The correlation coefficient is symmetric: $\gamma_{AB}(\tau)=\gamma_{BA}(\tau)$.

\section{\label{experiment}Experiment with Four Remote Clocks}
\begin{figure}[htbp]
\includegraphics[width=\linewidth]{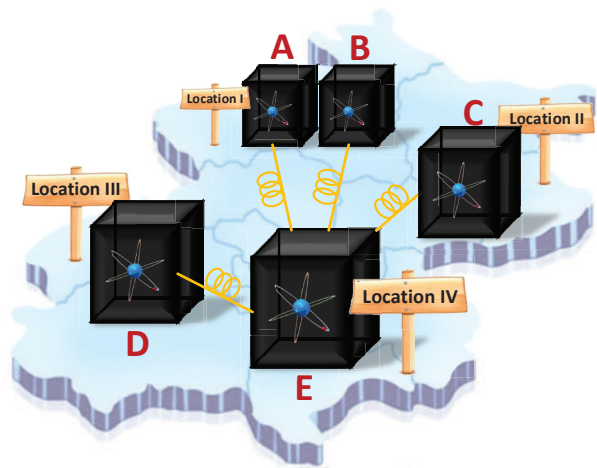}% Here is how to import EPS art
\caption{\label{fig1}Beijing regional frequency synchronization network. The letters A, B, C, D, and E denote various H-masers. The yellow solid lines indicate synchronization links using urban fibres.}
\end{figure}
It is necessary to include distributed clocks in the comparison for the correlation measurement as well as for other applications, such as to generate a time scale using a less-correlated clock ensemble. But the measurement of remote clocks was difficult to realize. Currently, the main method in use is via satellite links, for example, the technique of Global Positioning System (GPS) and two-way satellite time and frequency transfer (TWSTFT)~\cite{27}. These methods could achieve sub-nanosecond-level precision in real-time comparison and frequency stability at the 10-15/day level in terms of Allan deviation~\cite{28}. However, this is not sufficient to transfer signals of hydrogen masers at short averaging times~\cite{29}.
 
Since 2013, we have been working on organizing a Beijing regional frequency synchronization network~\cite{10, 21, 22}. As shown in Fig.~\ref{fig1}, using several homemade “frequency dissemination systems”, we connect a few time-keeping clocks from different institutes via urban fibre links. Clocks A, B, C, D, and E are all hydrogen masers, of which A and B are in the same room at location I, while all other clocks are located separately at different institutions. The lengths of the fibre links from location I, II and III to IV are 40 km, 23 km and 20 km, respectively. The 100-MHz frequency signals of clocks A, B, C, and D are recovered at location IV and can be simultaneously measured. The transfer stability of each fibre link is better than $5\times10^{-14}$/s and $5\times10^{-18}/10^6$s~\cite{10,30}, which can fully satisfy the frequency dissemination stability requirements of these H-masers~\cite{29}.

\begin{figure*}[htbp]
\begin{minipage}{\textwidth}
\includegraphics{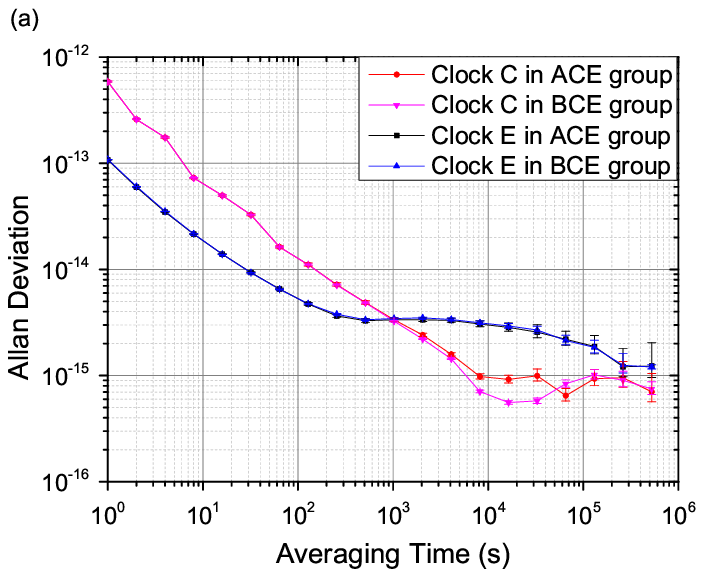}
\includegraphics{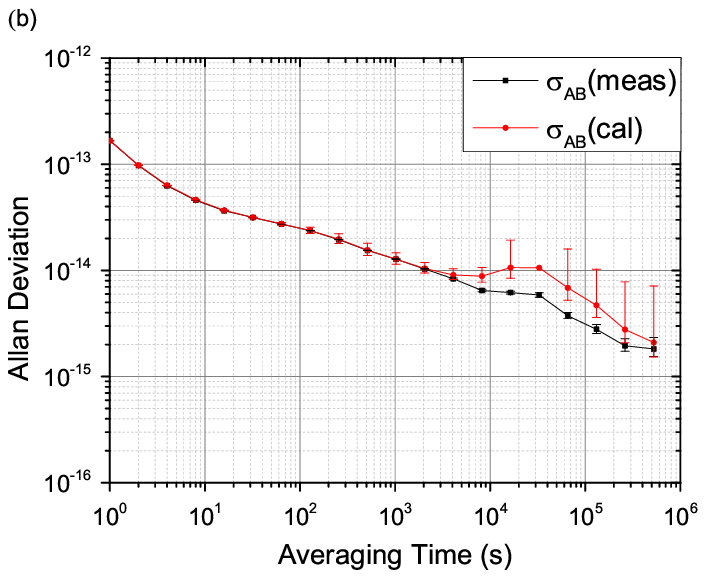}\\
\includegraphics[width=8.5cm]{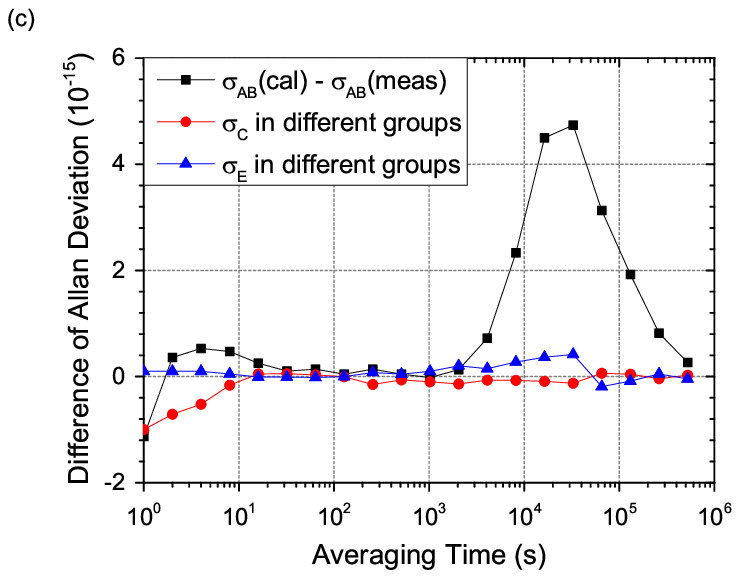}
\includegraphics[width=8.5cm]{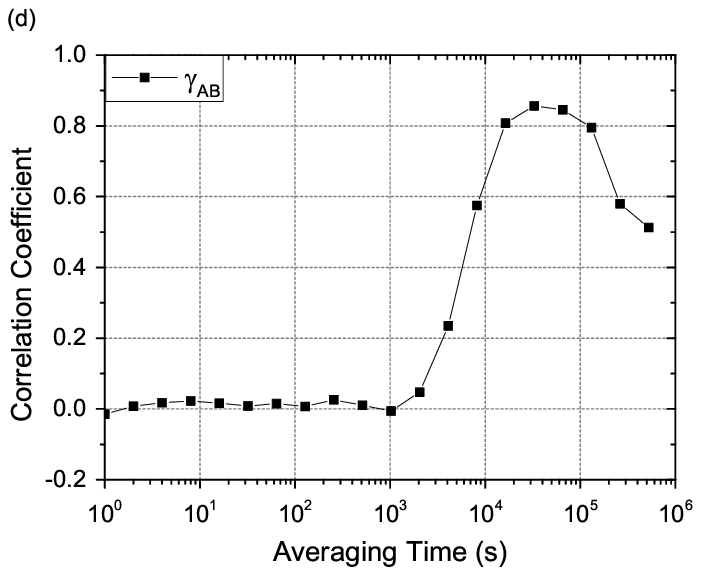}
\end{minipage}
\caption{Experimental results using H-masers A, B, C and E. (a) Individual frequency stabilities of H-masers C and E calculated using the 3-cornered-hat method in groups ACE and BCE. (b) Measured and calculated relative frequency stabilities between H-masers A and B. (c) The difference of $\sigma_C$ and $\sigma_E$ from different clock groups in (a) and the difference between $\sigma_{AB}(cal)$ and $\sigma_{AB}(meas)$ in (b). (d) Coefficient of clock correlation, $\gamma_{AB}(\tau)$, between H-masers A and B.}
\label{fig2}
\end{figure*}
A continuous measurement was performed several times utilizing this frequency synchronization network, using four H-masers each time. Fig.~\ref{fig2} shows the measurement results obtained over 59 days using H-masers A, B, C and E. We separate the clocks into two groups, ACE and BCE. In each group, the clocks are all at different locations. Therefore, the correlations among these clocks are negligible. Thus, we obtain the true values of the individual frequency stabilities by using the 3-cornered-hat method~\cite{23,24,25,26} for each group.

Because each group contains clocks C and E, the stabilities of clocks C and E will have two sets of results, which are shown in Fig.~\ref{fig2}(a). The error bars are given using the method introduced in~\cite{31}. The plot shows that the frequency stabilities of the same clock in different groups are well matched, indicating sufficiently high precision of the measurement.

In addition, we obtain the individual frequency stabilities of clocks A and B. Considering the ideal case in which there is no correlation between clocks A and B, their relative stability can be calculated as $\sigma^2_{AB}(cal)=\sigma^2_{A}+\sigma^2_{B}$. Fig.~\ref{fig2}(b) shows a comparison between the calculated relative stability $\sigma_{AB}(cal)$ and the measured relative stability $\sigma_{AB}(meas)$ for different averaging times. The difference between these values arises because $\sigma^2_{AB}(meas)$ contains the correlation term. Since both H-masers share the same environment, their correlation term is positive. Thus, the measured stability (black squares) is better than the calculated value (red circles). By comparing these curves, we determine not only the magnitude of the correlation term but also the averaging time at which this correlation appears.

In order to determine the effect of system noise on the correlation measurement, we plot, in Fig.~\ref{fig2}(c), the differences of the Allan deviations for a given clock in different groups based on Fig.~\ref{fig2}(a) as well as the difference between $\sigma_{AB}(cal)$ and $\sigma_{AB}(meas)$ in Fig~\ref{fig2}(b). Based on the comparison, it is clear that the error induced by the measurement and the algorithm, which is indicated by the differences of $\sigma_{C}$ and $\sigma_{E}$ from different groups, remains relatively low for all averaging times. Meanwhile, the difference $[\sigma_{AB}(cal)-\sigma_{AB}(meas)]$ induced by the correlation becomes significantly greater for averaging times ranging from $\sim10^3$s to $\sim10^5$s. This result indicates that the measurement system is sufficiently precise for the experiment within this period. Fig.~\ref{fig2}(d) shows the correlation coefficient of H-masers A and B, as defined in section~\ref{coefficient}. We see from this figure that $\gamma_{AB}(\tau)$ remains around zero for an averaging time of less than $\sim10^3$s. The value becomes positive and increases rapidly to more than 0.8 for averaging times ranging from $\sim10^3$s to $\sim10^5$s. Beyond that point, the correlation coefficient decreases to approximately 0.5 at an averaging time of 5 days.

\begin{figure*}[htbp]
\begin{minipage}{\textwidth}
\includegraphics{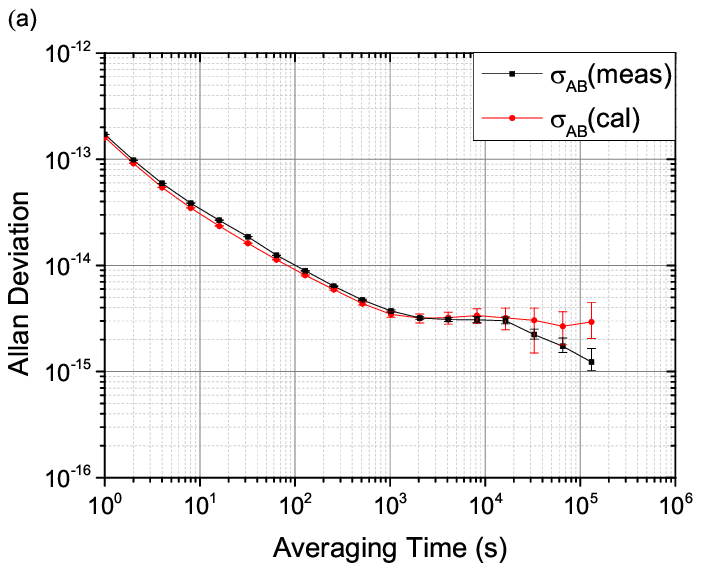}
\includegraphics{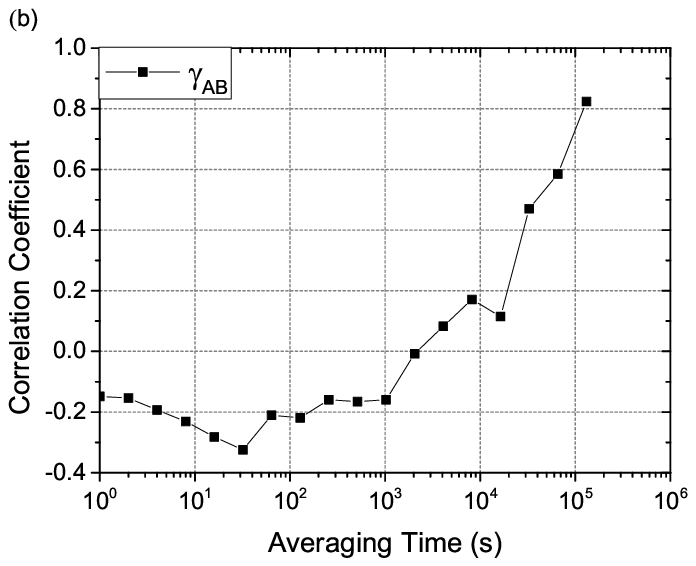}
\end{minipage}
\caption{Test results for H-masers A, B, D and E. (a) Measured and calculated relative frequency stabilities between H-masers A and B. (b) Coefficient of clock correlation, $\gamma_{AB}(\tau)$, between H-masers A and B.}
\label{fig3}
\end{figure*}
Using H-masers A, B, D, and E, we repeated the correlation measurement experiment. After 13 days of continuous measurement, we obtained the calculated and measured relative frequency stabilities of clocks A and B, as well as their correlation coefficient, as shown in Fig.~\ref{fig3}. The results of this test are in accordance with those of the first experiment. The correlation coefficient $\gamma_{AB}(\tau)$ increases beyond an averaging time of $\sim10^3$s  and reaches 0.8 at $\sim10^5$s.
\section{Simulation and Discussion}
To determine the cause of the correlation, we simulated 4 H-masers at 3 different locations and calculated the correlation between two H-masers with the same environmental factors. Using the method described in~\cite{32,33,34,35,36,37}, four H-masers are modeled as follows:
\begin{equation}
\begin{aligned}
x_i(t)={}& x_i(0)+y_i(0)\cdot t+\frac{1}{2}d_i\cdot t^2+\mu_i^1\cdot W_i^1(t)+{}\\
&\mu_i^2\cdot\int_0^tW_i^2(s)\cdot ds, i=1,2,3,4.
\end{aligned}
\label{eq13}
\end{equation}
Here, $x_i$ and $y_i$ are, respectively, the phase time and fractional frequency of clock i. The parameter $d_i$ represents the frequency drift or aging of clock i. $W_i^k(t),k=1,2$, are two independent, one-dimensional standard Wiener processes (standard Brownian motion). The two terms containing these processes represent different types of noise. $\mu_i^1\cdot W_i^1(t)$ represents white frequency modulation noise (WFM), and $\mu_i^2\cdot\int_0^tW_i^2(s)\cdot ds$ represents random walk frequency modulation noise (RWFM). The constants $\mu_i^1$ and $\mu_i^2$ are diffusion coefficients and provide the intensity of each noise type.

To simplify the simulation, we set the same parameters for all four H-masers according to previous research on clock noise~\cite{3,35} and our analysis of the experimental H-masers given in section~\ref{experiment}:
\begin{equation}
\begin{aligned}
x_i(0){}& =0{}\\
y_i(0){}& =1\times10^{-12}{}\\
d_i{}& =0{}\\
\mu_i^1{}& =8.8\times10^{-14}{}\\
\mu_i^2{}& =5.6\times10^{-18},  i=1,2,3,4.{}
\end{aligned}
\label{eq14}
\end{equation}
In this way, we generate the original phase time data of 4 H-masers for 60 days with a time interval of 1s, which forms a matrix of $5184000\times4$. These clocks are ideal H-masers with no correlations among each other. As previously discussed, we believe that the correlation between clocks is mainly induced by common environmental factors such as temperature, pressure, relative humidity, magnetic fields variation, and so on. Experiments have shown that among these factors, temperature, magnetic fields and relative humidity have a major impact on the H-maser frequency~\cite{1}. Thus, we consider only these three factors in simulating the H-masers at different locations.

Both magnetic field and relative humidity changes have linear effects on the H-maser frequency change. While for temperature, both static and dynamic effects exist~\cite{1}. The frequency change is proportional to not only the change in temperature but also the rate of change in temperature. Different H-masers have different sensitivities to a given environmental factor. However, here, we use the same value of sensitivity for all H-masers in our simulation, as described in Table~\ref{t1}.

\begin{table}[htbp]
\caption{\label{t1}
Typical value of H-maser environmental sensitivities~\cite{1}
}
\begin{ruledtabular}
\begin{tabular}{lcd}
\textrm{Environmental parameter}&
\textrm{Sensitivity}&
\textrm{Typical value}\\
\hline
Static temperature & $S_{ST}$ & -5\times10^{-15}/^{\circ}C\\
Dynamic temperature & $S_{DT}$ &  -1\times10^{-14}/(^{\circ}C\cdot s^{-1}) \\
Magnetic field\footnote{Only the vertical magnetic field is considered because this is the most sensitive axis of an H-maser.} & $S_M$ & +8\times10^{-16}/\mu T \\
Relative humidity & $S_H$ & +2\times10^{-16}/\%\\
\end{tabular}
\end{ruledtabular}
\end{table}
Based on the environmental factors of a typical H-maser room, we create 3 sets of uncorrelated data for temperature, magnetic field and relative humidity, each for 60 days with a 1-second time interval.

Then, the four H-masers in three different locations can be modeled as follows:
\begin{equation}
\begin{aligned}
Y_i(t)={}&\frac{x_i(t+\tau)-x_i(t)}{\tau}+S_{ST}\cdot[(T_i(t)-T_i(0)]+{}\\
{}&S_{DT}\cdot\frac{dT_i(t)}{dt}+S_M\cdot[(M_i(t)-M_i(0)]+{}\\
&S_H\cdot[H_i(t)-H_i(0)], i=1,2,3,4.
\end{aligned}
\label{eq15}
\end{equation}
Here, $Y_i$ is the fractional frequency of maser i with environmental influences, and $x_i$ is the original phase time of maser i, as in Eq.~(\ref{eq13}). $\tau$ is the time interval and is set to 1s. $T_i$, $M_i$ and $H_i$ are the temperature, magnetic field and relative humidity of the ambient environment of maser i, respectively. We assume that H-masers 1 and 2 are in the same environment. Thus, they use the same set of environmental data, while H-masers 3 and 4 use two other sets of environmental data.

\begin{figure*}[htbp]
\begin{minipage}{\textwidth}
\includegraphics{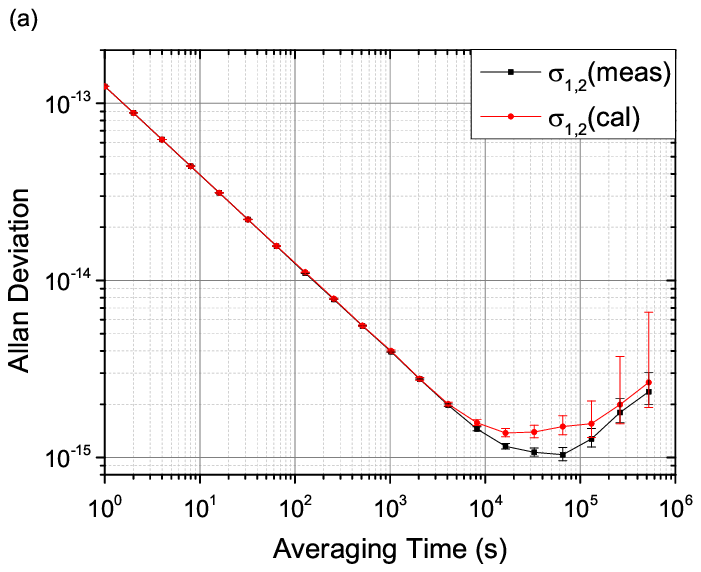}
\includegraphics{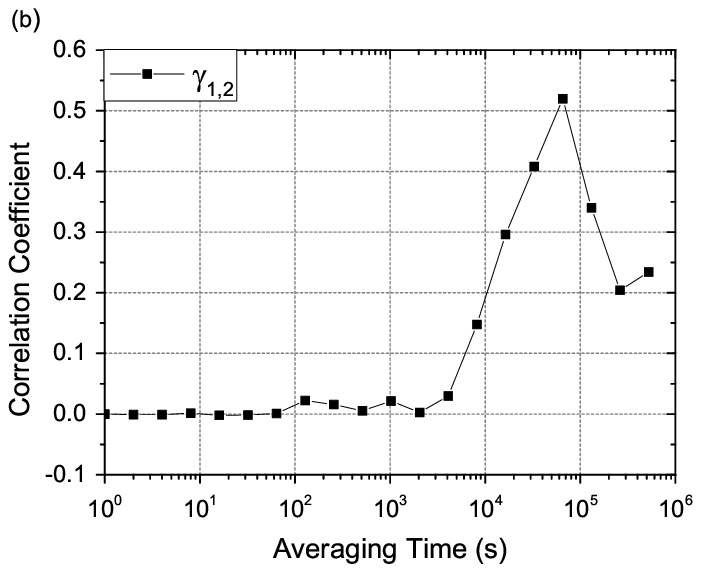}
\end{minipage}
\caption{Simulated results of 4 masers, of which H-masers 1 and 2 are at the same locations, while H-masers 3 and 4 are at other two different locations. (a) Measured and calculated relative frequency stabilities between H-masers 1 and 2. (b) Coefficient of clock correlation, $\gamma_{1,2}(\tau)$, between H-masers 1 and 2.}
\label{fig4}
\end{figure*}

Using these data, we analyse the correlation following the same approach utilized for the experiments. In Fig.~\ref{fig4}(a), the red circle and black square lines represent the calculated and measured relative frequency stabilities of H-masers 1 and 2. Fig.~\ref{fig4}(b) shows the correlation coefficient correspondingly. The coefficient is near zero for an averaging time of less than $\sim10^3$s, then becomes positive and grows rapidly to about 0.5 at approximately $\sim10^5$s, and decreases beyond $\sim10^5$s. The result of simulation is consistent with that of the experiments described in section~\ref{experiment}, demonstrating the precision of the transferring and measuring system in our tests and supporting the theory that the common environmental influence is the primary reason for clock correlation.

An analysis was performed to explore the causes for these properties of correlation. In a typical temperature-controlled room, the peak-to-peak values of temperature, the rate of temperature change, the magnetic field and the relative humidity fluctuations are approximately 1$^{\circ}C$, 0.05$^{\circ}C/$s, 0.12$\mu T$ and 3.5\%, respectively, which leads to a maximum H-maser frequency change at the magnitude of $10^{-15}$, according to the sensitivities listed in Table~\ref{t1}. From both the experimental and simulative results, we found that the relative frequency instability (Allan deviation) of the two H-masers is too large to exhibit a frequency change at this level before $\sim10^{3}$s. For this reason, the correlation is not observed until the averaging time increases to $\sim10^{3}$s. In our H-maser model, the diffusion coefficient of the RWFM noise is set as $\mu_i^2=5.6\times10^{-18}$, which is calculated from the obervation of H-masers in the experiment. Since RWFM noise grows at a speed of $\tau^{\frac{1}{2}}$ in terms of the Allan deviation, at an averaging time of $\sim10^5$s, the noise will reach the level of $10^{-15}$ and continue to increase. Therefore, beyond $\sim10^5$s, the RWFM noise of the H-maser itself is larger than the noise induced by the environment, so the correlation is “covered” by the RWFM noise. This leads to the decrease of correlation coefficient in this period as observed in both experiments and simulations, because the Allan deviations, $\sigma_1$ and $\sigma_2$, who are the dominators, increase.
\section{\label{conclusion}Conclusion}
In summary, we defined the correlation coefficient of two atomic clocks and measured the correlation between two co-located H-masers using a high-precision fibre-based frequency synchronization network in Beijing. Furthermore, we built an H-maser model with environmental influences and calculated the correlation between two H-masers in the same environment. The results of the experiment and simulation are in good agreement with each other. The correlation coefficient between co-located H-masers remains small for averaging times of less than $\sim10^{3}$s, then becomes positive and increases over the range of $\sim10^{3}$s to $\sim10^{5}$s, and decreases beyond $\sim10^{5}$s. The correlation coefficient can be as high as 0.8, and its consequence in long-term, precise time-keeping using co-located clock assembly should be carefully considered.
\section{Acknowledgement}
This work was supported by National Key Project of Research and Development (No.2016YFA0302102) and the Program of International S\&T Cooperation (No.2016YFE0100200).

% The \nocite command causes all entries in a bibliography to be printed out
% whether or not they are actually referenced in the text. This is appropriate
% for the sample file to show the different styles of references, but authors
% most likely will not want to use it.
\nocite{*}

%\bibliography{apssamp}% Produces the bibliography via BibTeX.

\end{document}